\documentclass[3p,times,twocolumn]{elsarticle}

\usepackage{amssymb}

\usepackage[figuresright]{rotating}

\begin{document}

\begin{frontmatter}




\title{Search for time dependence of the $^{137}$Cs decay constant}


\author[mi]{E. Bellotti}
\author[pd]{C. Broggini}
\author[lngs]{G. Di Carlo}
\author[lngs]{M. Laubenstein}
\author[pd]{R. Menegazzo}

\address[mi]{Universit\`{a} degli Studi di Milano Bicocca and Istituto Nazionale di Fisica Nucleare, Sezione di Milano, Milano, Italy}
\address[pd]{Istituto Nazionale di Fisica Nucleare, Sezione di Padova, Padova, Italy}
\address[lngs]{Istituto Nazionale di Fisica Nucleare, Laboratori Nazionali del Gran Sasso, Assergi (AQ), Italy}

\begin{abstract}
Starting from June 2011, the activity of a $^{137}$Cs source has been measured by means of a HPGe detector installed deep underground in the Gran Sasso Laboratory. 
In total about 5100 energy spectra, one hour measuring time each, have been collected. These data allowed the search for time variations of the 
decay constant with periods from a few hours to 1 year. No signal with amplitude larger than 9.6$\cdot$10$^{-5}$ at 95$\%$ C.L. has been detected. These limits are more than one order of magnitude lower than the values on the 
oscillation amplitude reported in literature. In particular, for 1 year period an oscillation amplitude larger than 
8.5$\cdot$10$^{-5}$ has been excluded at 95$\%$ C.L., independently of the phase. The same data give a value of  29.96$\pm$0.08 years  
for the $^{137}$Cs half life, in good agreement with the world mean value of  30.05$\pm$0.08 years.
\end{abstract}

\end{frontmatter}


\vspace*{0.5cm}

\section{Introduction}
\label{}
The possible time dependence of the radioactive nuclei decay constant has been investigated since the infancy of the science of radioactivity \cite{Cur03}.
Recently, the interest in the time dependence has strongly increased \cite{Jen09} since
various experiments have reported evidence of such an effect.
In particular, in \cite{Sie98} the constancy of the activity of a $^{226}$Ra source has been measured with an ionization chamber. 
An annual modulation of amplitude 0.15$\%$, having the maximum in February and the minimum in August, has been observed. The activity of a $^{152}$Eu source has also been measured by means of a Ge(Li) detector and an even larger 
annual modulation (0.5$\%$) has been detected in the counting rate at the full energy peak of the 1408 keV  line.
Alburger et al. \cite{Alb86}  measured the half-life of $^{32}$Si, which is interesting for many applications in Earth Science. Data, collected in four years, show an evident annual modulation, 
with an amplitude of about 0.1$\%$ .
In \cite{Jen09,Fis11} the existence of new and unknown particle interaction has been advocated to explain the yearly variation in the activity of radioactive sources \cite{Sie98,Alb86}.
The Authors correlate these variations, whose  
amplitude is about 10$^{-3}$, the maximum being around February and the minimum in August, to the Sun-Earth distance.

The paper by Jenkins and Collaborators \cite{Jen09} triggered a strong interest in the subject. Old and recent data have been analyzed, or reanalyzed, to search for periodic and sporadic variations with time. 
Some of the measurements and analysis confirm the existence of oscillations \cite{Par10,Jav10} whereas others contradict this hypothesis \cite{Har11,Coo08,Nor08}.
For instance, \cite{Par10} presents the time dependence of  the counting rate  for $^{60}$Co, $^{90}$Sr and $^{90}$Y sources, measured with Geiger M\"{u}ller detectors and for a $^{239}$Pu source, 
measured with silicon detectors. 
While beta sources show an annual (and monthly) variation with amplitude of about 0.3$\%$, the count rate from the Pu source is fairly constant. 

More recently, an anti-correlation between decay rate and intensity of solar flares has been suggested. As a matter of fact, in \cite{Jen06} the decay rate of  a $^{54}$Mn source has been measured and a significant 
dip in the count rate in coincidence with solar flares has been observed. According to the Authors, this suggests that nuclear decay rates may be affected by solar activity. 
However, in \cite{Par06} no anti-correlation is found between  decay rate and intensity of solar flares. 

As also suggested by many authors, we believe that more dedicated experiments are needed to clarify if the observed effects are physical or are due to systematic effects not taken into account.
 Therefore, we performed a dedicated experiment using a $^{137}$Cs source. Time dependence of the order of 0.2$\%$ has already been reported \cite{Bau01} in the decay constant of $^{137}$Cs measured with a germanium detector.

The special feature of our experiment is the set-up installed deep underground in the INFN Gran Sasso Underground Laboratory. The laboratory conditions are very favorable: the cosmic ray 
flux is reduced by a factor 10$^{6}$ and the neutron flux by a factor 10$^{3}$ with respect to above ground.  As a consequence, we do not have to take into account possible 
time variations of these fluxes since their contribution to the counting rate is completely negligible. Moreover, the laboratory temperature is naturally constant, maximum variation 
being of a few degrees Celsius during the year.

\section{The measurement}
\label{}
The aim of the present experiment is the measurement of a $^{137}$Cs source activity as a function of time by the detection and the counting of the
emitted gamma rays.

\subsection{The set-up}
The set-up (Fig. 1) is installed in the low background facility STELLA (SubTErranean Low Level Assay) located in the underground laboratories of LNGS. 

\begin{figure}[t] 
   \centering
   \includegraphics[width=2.8in]{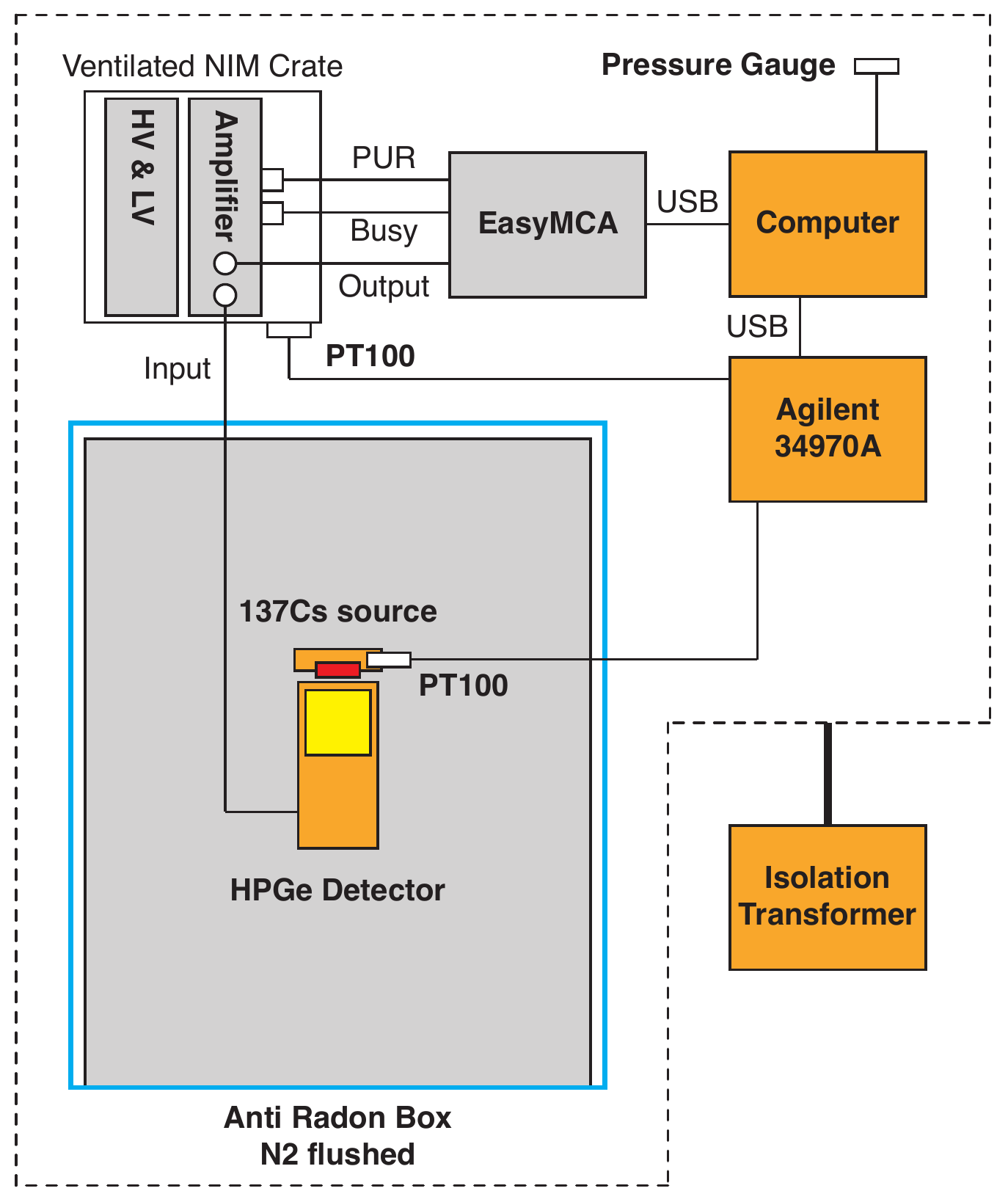} 
   \caption{Layout of the setup. The whole system is powered through an  isolation transformer.}
   \label{fig:setup}
\end{figure}

The source is a $^{137}$Cs standard source with Cs embedded in a plastic disk of 1" diameter and 1/8" thick. The source nominal activity was 4.14 kBq (July 16$^{th}$, 1998). 
At the beginning of the measurement (June 6$^{th}$, 2011) its activity has been estimated to be 3.0 kBq. 

The detector is a p-type High Purity Germanium manufactured by Ortec and powered at the nominal bias of 3500 V. The crystal has 78.4 mm diameter and is 84.6 mm high with a relative efficiency of 96$\%$. 
The source is firmly fixed to the copper end-cap of the germanium detector in order to minimize variations in the source-detector relative positions. As a matter of fact, from Monte Carlo simulations we have
that a variation of 1 micron in the source-detector distance would cause a variation of 5$\cdot$10$^{-5}$ in the counting efficiency. 

\begin{figure*}[htb] 
   \centering
   \includegraphics[width=6.20in]{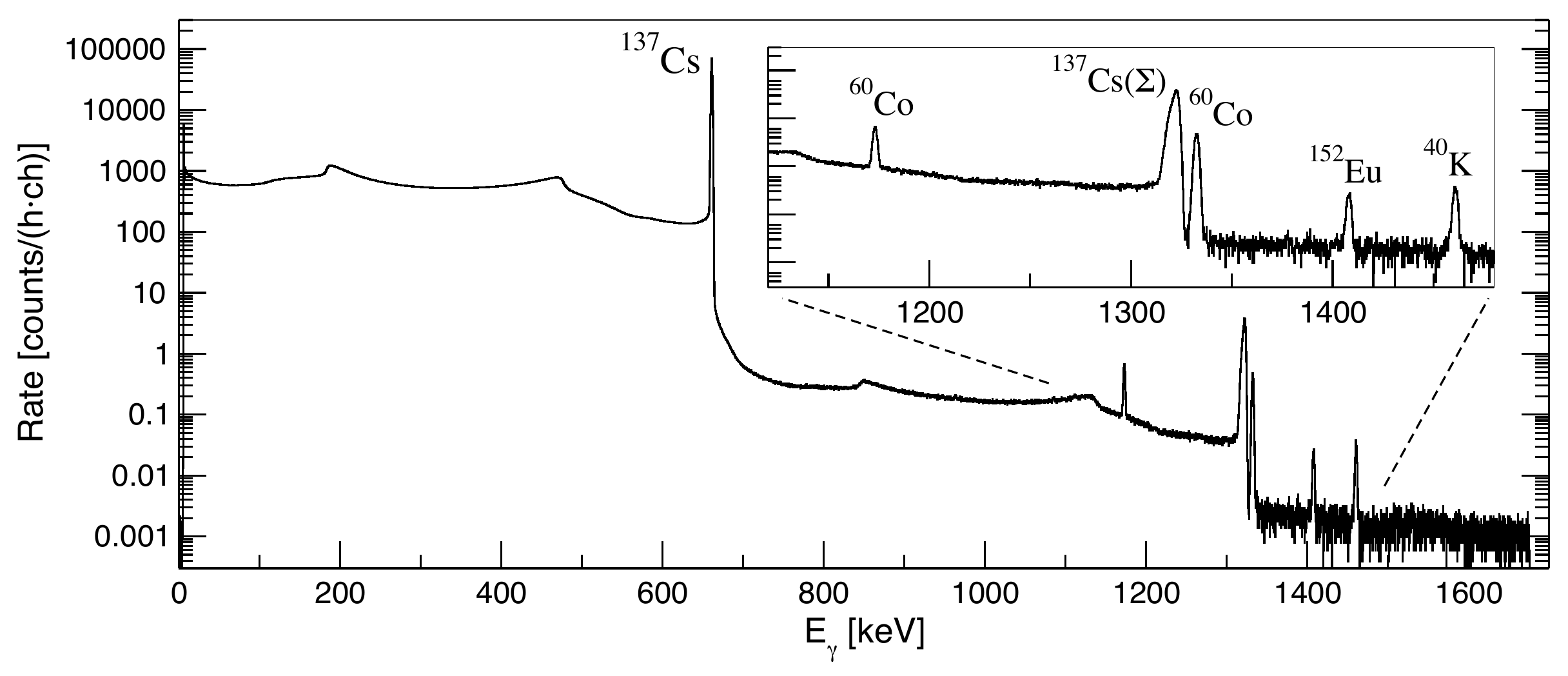} 
   \caption{Measured $\gamma$-ray spectrum of the $^{137}$Cs source after 5124 hours. A weak activity from $^{60}$Co, $^{152}$Eu and $^{40}$K contaminants is observed at high energy (see inset).
    The $^{40}$K line is also evident in the background spectrum. The peak at 1.323 MeV results from the accidental coincidence of two 661.6 keV $\gamma$-lines when two $^{137}$Cs decays occur within the pulse pair 
    resolution of the spectroscopy amplifier.}
   \label{fig:spe1}
\end{figure*}

The Germanium detector is surrounded by at least 5 cm of
copper followed by 25 cm of lead to suppress the laboratory gamma ray background. Finally, shielding and detector are housed in a polymethylmetacrilate box flushed with nitrogen at slight overpressure and which is 
working as an 
anti-radon shield. 

The signal from the detector pre-amplifier goes first to an Ortec amplifier (mod.120-5F) where it is shaped with 6 $\mu$s shaping time, and then to a Multi Channel Analyser (Easy-MCA 8k Ortec, 0.21 keV/channel). Also 
the busy and pile-up signals from the amplifier are sent to the MCA. The pile-up rejection circuit of the amplifier is able to recognize two signals if they are separated in time by at least 0.5 $\mu$s. In order to minimize 
the noise the whole set-up, detector and electronics modules, is powered through a 3 kVA AC-AC isolation transformer. 

Finally, temperature and atmospheric pressure are the two important parameters we have to monitor since they could change the detector efficiency by varying the relative distance Cs source-detector. In particular, 
the temperature is continuously measured with a Pt100 sensor located in the shielding very close to the Cs source, whereas the atmospheric pressure is measured in a 
nearby hall of the underground laboratory. The temperature at the position of the electronics modules is also continuously monitored since it can affect the electronics performance.

\subsection{Data taking}
Spectra from the MCA are collected every hour (real time provided by the MCA), except during the liquid nitrogen refilling of the detector. These interruptions occur two times a week and last less than 2 hours. 
The quality of the measurement is monitored by checking the energy resolution at the Cs line (661.6 keV). Its value, 1.79$\pm$0.02 keV, is stable, this way proving  
that the electronic noise does not change with time. This is also confirmed by the counting rate in the lowest MCA channels, below 7 keV,
where the noise is dominant and whose rate remains constant with time. 

\begin{figure}[ht] 
   \centering
   \includegraphics[width=3in]{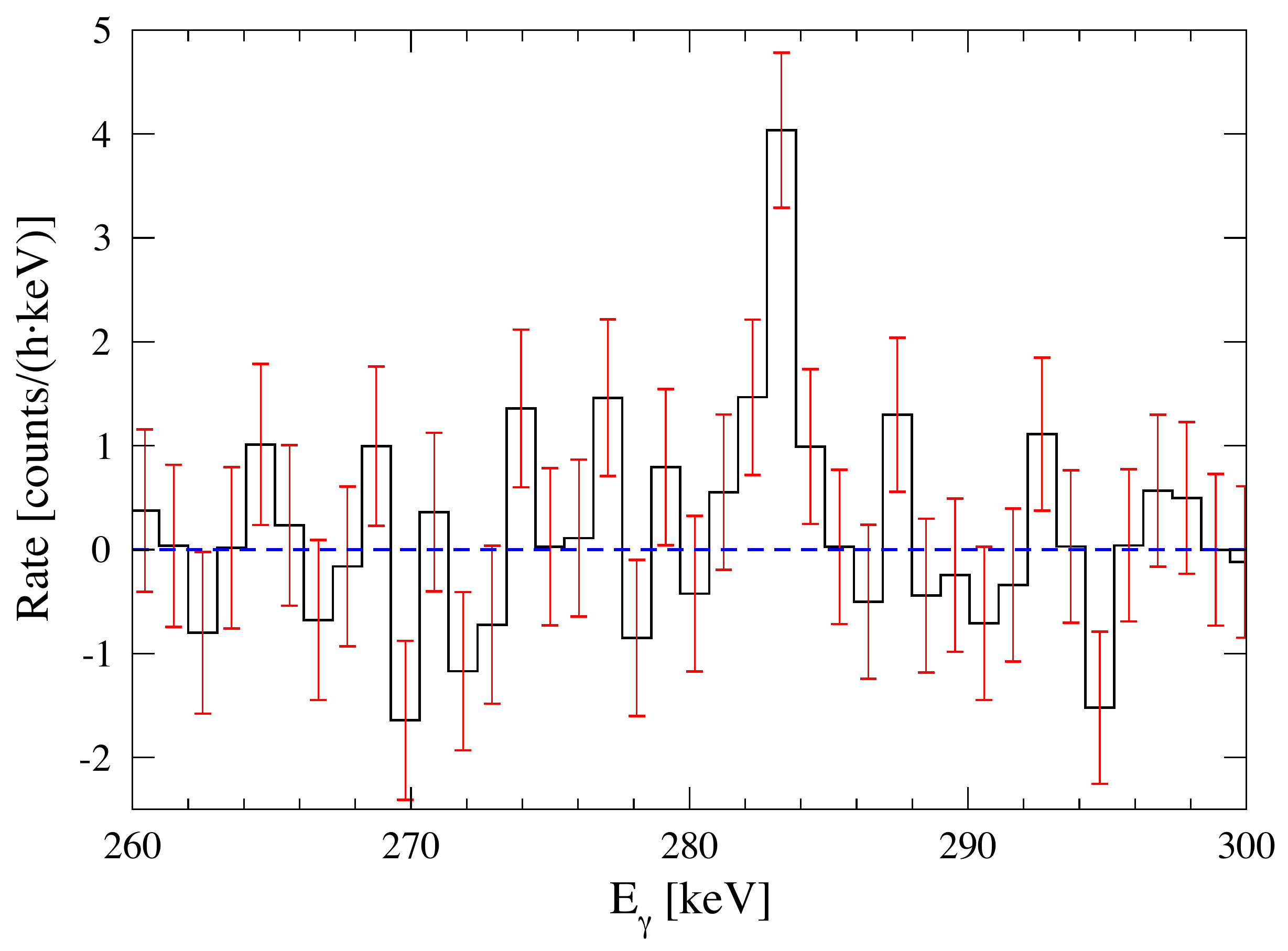} 
   \caption{Weak branching in the $^{137}$Cs spectrum from the decay of the (1/2)$^{+}$ level in $^{137}$Ba. The spectrum is obtained after a background subtraction. The background  component has been evaluated 
   from a polynomial fit of regions on both sides of the peak. Statistical uncertainties are indicated.}
   \label{fig:spe2}
\end{figure} 

\begin{figure*}[t] 
   \centering
   \includegraphics[width=\textwidth]{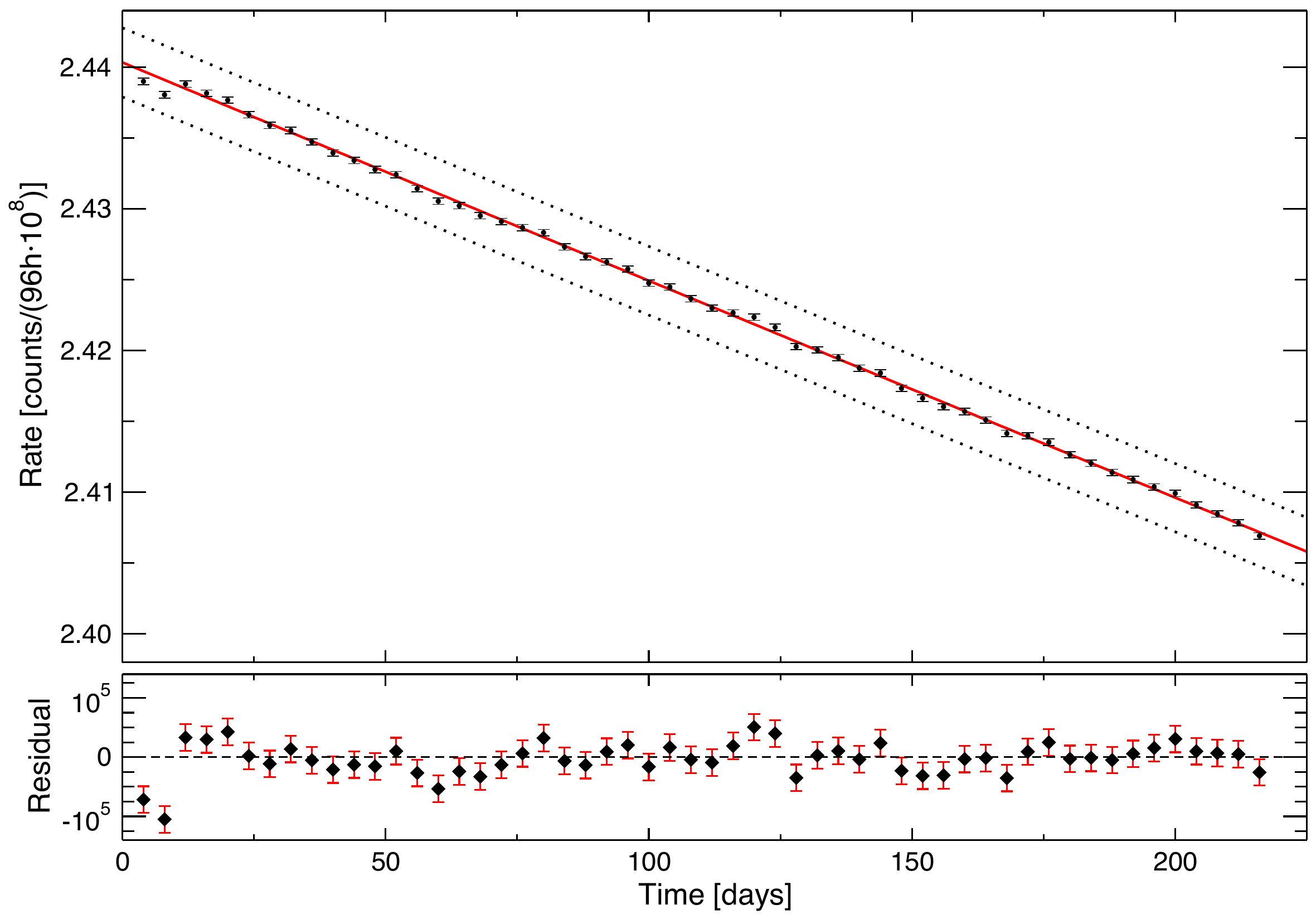} 
   \caption{Detected activity of the $^{137}$Cs source. Dead-time corrected data are summed over 96 hours. The first two points correspond to the beginning of data taking,
    when the set-up was stabilizing, and they are not considered in the analysis. Dotted lines represent a 0.1$\%$ deviation from the exponential trend. Residuals (lower panel) of the measured activity
    to the exponential fit. Error bars include statistical uncertainties and fluctuations in the measured dead time.}
   \label{fig:act}
\end{figure*}

The position of the Cs peak
is also constant, within 0.21 keV, proving the stability of the electronics. Finally, we monitored the dead time, provided by the MCA. It smoothly changed from the initial value of 5.10$\%$ to 4.98$\%$ as a 
consequence of the decreased 
activity of the Cs source.

\subsection{Energy spectrum}
The spectrum measured with the present set-up is shown in Fig. 2. 
Modest signals from $^{40}$K, $^{60}$Co and $^{152}$Eu are visible in the energy spectrum above 1 MeV. 
The $^{40}$K line is also present in the background spectrum, while the 
$^{60}$Co and $^{152}$Eu activities are related to the source. Their contribution to the total count rate, estimated by Monte Carlo simulation,
is 1.7$\cdot$10$^{-2}$ Hz for $^{60}$Co and 2.7$\cdot$10$^{-3}$ for $^{152}$Eu and $^{40}$K. The total count rate above the threshold of 7 keV is of about 700 Hz. 
The intrinsic background, i.e shielded detector without Cs source, has been measured during a period of 70 days. 
Thanks to the underground environment and to the detector shielding, it is very low, down to about 40 counts/hour above the 7 keV threshold (0.01 Hz).

The signature of the $^{137}$Cs source is given by the outstanding peak at 
661.6 keV, by the peak at 1.323 MeV due to the sum of two 661.6 keV gamma rays too close in time to be recognized by the pile-up circuit and, finally, by the modest peak at 283 keV (Fig. 3). 
This last peak is due to the (1/2)$^{+}$ level of $^{137}$Ba populated by the beta decay of $^{137}$Cs with a branching ratio of 5.3(14)$\cdot$10$^{-6}$ \cite{Bik96}.

\section{Data analysis}
\label{}
Data discussed in this letter have been continuously collected during 217 days, starting from June 6th 2011 to January 9th 2012. The activity of the source can be calculated for any interval of time multiple of the hour.
In particular, Fig. 4 shows the activity per 4 days as a function of time, i.e. the integral from 7 keV to 1.7 MeV of the MCA spectra collected during 4 days and corrected for the dead time.
A total error equal to the linear sum of the statistical uncertainty (6.4$\cdot$10$^{-5}$ relative) and of the fluctuations of the dead time (2.8$\cdot$10$^{-5}$ relative) has been assigned to each data point.
The activity decrease is due to the source decay. The continuous line in Fig.4 is the exponential fit obtained with the mean life which minimizes the Chi Squared: 43.22 years
(we take the year of 365.25 days of 86400 seconds each). The reduced Chi Squared per degree of freedom is 1.02. 
During the 7 month period of data taking the temperature at the Cs source smoothly increased by 0.4 K with a few larger variations up to 0.7 K during one week. There is no effects on the data 
correlated with such temperature variations.
In the same period the pressure inside the laboratory varies in the range $\pm$10 mbar. Also for these pressure changes we can exclude any significant effect on the rate.

\subsection{$^{137}$Cs half-life}
The $^{137}$Cs half-life has been determined by many authors during more than fifty years (for a critical review of available data see \cite{Hel11}). Data collected in this measurement 
allow for a new and precise estimate of the 
$^{137}$Cs half-life.
The data are  fitted with an exponential function leaving two parameters free: initial decay rate and half-life. 
The resulting half-life is 10942$\pm$30 days, to be compared with the recommended value of 10976$\pm$30 days.

\subsection{Time modulations}
We searched for time variations with periods from 6 hours to 400 days. For short periods, up to 40 days,
the use of the discrete Fourier transform of the hourly data, after the subtraction of the exponential trend, is appropriate.
For longer period we performed a Chi Squared analysis of the daily count rate, fitting with a superposition of the exponential decay function (with our mean life time estimate 
of 43.22 years) and a sine function of fixed period. The initial activity, the amplitude of the oscillation and its phase are left free. We scanned all the periods from 40 to 400 days, studying in particular
the 1 solar year period. No significant improvement of the Chi Squared per degree of freedom has been observed in this range. Our analysis excludes any oscillation with amplitude larger  
 than 9.6$\cdot$10$^{-5}$ at 95$\%$ C.L.. In particular, for an oscillation period of 1 year the amplitude is 3.1(2.7)$\cdot$10$^{-5}$, well compatible with zero; and a limit of 8.5$\cdot$10$^{-5}$ at 95$\%$ C.L.
on the maximum allowed amplitude is set independently of the phase. 
As a consequence, the claimed existence of an oscillation with amplitude of about 10$^{-3}$ due to the variation of the Sun to Earth distance is rejected by the present data.

\section{Conclusion}
\label{}
The half-life of a $^{137}$Cs source has been estimated to be 10942$\pm$30 days, in agreement with its recommended value.
Moreover, from our measurement we can exclude the presence of time variations in the source activity, superimposed to the expected exponential decay, larger than 9.6$\cdot$10$^{-5}$ at 95$\%$ C.L. for 
oscillation periods in the range 6 hours-1 year.
In particular, we exclude an oscillation amplitude larger than 8.5$\cdot$10$^{-5}$ at 95$\%$ C.L. correlated to the variation of the Sun-Earth distance,
in clear contradiction with previous experimental results and their interpretation as indication of a novel field (or particle) from the Sun.
The data taking with the improved experimental set-up is going to continue for at least another 6 months.

\section {Acknowledgments}
Thanks are due to our colleagues C. Cattadori, L. Pandola and C. Rossi Alvarez for many useful discussions; to M. Dedeo who provided us data on atmospheric pressure, to
L. Votano, director of the LNGS, for the constructive hospitality.








\end{document}